\documentclass[a4paper]{jpconf}
\usepackage{graphicx}
\usepackage{amsmath,amssymb}
\begin{document}
\title{Can Spectral Action be a Window to Very High Energies?}

\author{Dmitri Vassilevich}

\address{CMCC, Universiade Federal do ABC, Santo Andr\'e, SP, Brazil}

\ead{dvassil@gmail.com}

\begin{abstract}
The principles of noncommutative geometry impose severe restrictions on the structure of (almost)
commutative field theories. The Standard Model fits surprisingly well into the noncommutative framework. Here we overview some universal predictions of the spectral action principle for the
behavior of bosonic theories at very high energies.
\end{abstract}

\section{Introduction}
Nowadays, no one doubts that quantum theory is an adequate instrument to describe the real world. There are strong theoretical arguments that the space-time itself can be quantum. Noncommutative Geometry provides a consistent mathematical framework to deal with quantum spaces. This framework appears to be quite restrictive, so that it yields testable predictions even on commutative spaces. In particular, the principles of Noncommutative Geometry have very interesting consequences for particle physics and gravity that will be the subject of present contribution.

For the readers' convenience we start with some basic references. A fairly complete introduction to Noncommutative Geometry is the monograph \cite{GVF}. More recent developments including the applications to particle physics can be found in \cite{CM} and \cite{WalterB}. 

Geometry of noncommutative spaces is defined through Spectral Triples $(\mathcal{A},\mathcal{H},\mathcal{D})$. Here $\mathcal{A}$ is an associative algebra that acts by bounded operators on some Hilbert space $\mathcal{H}$, while $\mathcal{D}$ is a Dirac operator on $\mathcal{H}$. The first two entries of Spectral Triple remind us of the Gelfand-Naimark theory. Any commutative $C^*$ algebra is isomorphic to the algebra of continuous functions on some topological space. In this sense, a noncommutative associative $C^*$ algebra can be thought of as the algebra of functions on a noncommutative space. Abstract algebras are hard to deal with, one has to realize them on some space. The Gelfand-Naimark-Segal construction provides any $C^*$ algebra with a Hilbert space where this algebra acts by bounded operators. Thus, $\mathcal{A}$ and 
$\mathcal{H}$ serve to describe topology of (possibly) noncommutative spaces. However, just $\mathcal{A}$ and $\mathcal{H}$ are not enough to describe geometries. One needs a scale to measure the distances, which is supplied by the third member of Connes' Spectral Triple -- the Dirac operator $\mathcal{D}$. Spectral Triples should satisfy a number of Axioms that we do not write down here. 

The Dirac operator may contain a lot of important data as, e.g., the Riemann metric, connections, etc. It is natural to assume that fundamental
physics is a geometric theory, i.e., it is fully based on the Spectral Triple. One needs therefore to construct an action principle basing on the structures available in the Spectral Triple only. This construction is called the Spectral Action principle \cite{CC}. The action for spinor fields that belong to the Hilbert space $\mathcal{H}$ may be constructed with the help of inner product $\langle \ ,\ \rangle$ and the Dirac operator
\begin{equation}
\mathcal{S}_{\rm f}=\langle \psi, \mathcal{D}\psi \rangle \label{Sf}
\end{equation}
This action has the form of the usual fermionic action. The part of the action that does not contain fermions should be constructed from 
$\mathcal{D}$ only. To get a number out of an operator one can take the trace, so that the bosonic part of the spectral action becomes
\begin{equation}
\mathcal{S}_{\bf b}={\rm Tr}_{\mathcal{H}}\, \big( \chi (\mathcal{D}^2/\Lambda^2) \big) \label{Sb}
\end{equation}
Here $\chi$ is any function that decays fast enough at infinity to ensure the existence of trace. $\Lambda$ is a scale parameter. 

Spectral triples also include the operators of chirality and real structure. These operators  will not be considered in this contribution even
though they play an important role.

Apart from the Spectral Action, spectral triples allow to define and compute many other important characteristics of noncommutative spaces (as the distance, e.g.). In particular, any abstract spectral triple with a commutative algebra $\mathcal{A}$ allows to reconstruct the corresponding commutative manifold by what is called the Connes Reconstruction Theorem. 
More generally, one may be interested in how much the spectrum of natural operators tells about the geometry. This is one of the subjects of Spectral Geometry, that has many applications to Quantum Field Theory \cite{FV}. Many notions and methods of spectral geometry may be extended even to differential operators on infinite rank bundles \cite{SympDir}.

\section{Low and moderate energies}
Let us see how the Standard Model of elementary particles looks from the point of view of Noncommutative Geometry.
Following the approach of \cite{SM1}, let us take $\mathcal{A}=C^\infty (\mathbb{R}^4)\otimes \mathcal{A}_{\rm SM}$, where 
$\mathcal{A}_{\rm SM}=\mathbb{C}\oplus \mathbb{H} \oplus {\rm Mat}_3(\mathbb{C})$ is finite dimensional. Here $\mathbb{H}$ 
is the algebra of quaternions. The corresponding noncommutative geometry is called almost commutative since the whole noncommutativity 
is in the finite-dimensional algebra $\mathcal{A}_{\rm SM}$. This finite-dimensional piece defines the gauge group which appears to be 
that of the Standard Model. The Hilbert space $\mathcal{H}$ consists of square integrable spinors that carry a representation of 
the associative algebra $\mathcal{A}_{\rm SM}$. There are not that many representations of associative algebras (as compared to the Lie algebras).
Therefore, the last condition restricts the spectrum of fermions essentially to trivial or fundamental representations of the $SU(2)$ and 
$SU(3)$ factors in the gauge group. Although the gauge group itself is defined by $\mathcal{A}_{\rm SM}$ that and may vary in particular realizations of almost commutative geometries, the spectrum of fermionic representations is fixed to the one that we do see in the Standard Model. Note, that the Higgs field appears very natural in this approach as fluctuation of the connection in internal noncommutative directions.

Let us see how much one could say about the dynamics. The fermionic action of the Standard Model defines through (\ref{Sf}) the Dirac
operator (almost) uniquely. The Dirac operator defines in turn the bosonic action (\ref{Sb}) that depends also on an arbitrary function
$\chi$. It happens, however, that in the low-energy limit the whole dependence of $\chi$ may be reduced to a few constants. To demonstrate this,
we need the heat kernel expansion for $\mathcal{D}^2$:
\begin{equation}
K(\mathcal{D}^2,t)\equiv {\rm Tr}\, \left[ e^{-t\mathcal{D}^2}\right] \simeq \sum_{p=0}^\infty
t^{-2+p} a_{2p}\bigl( D^2\bigr)\,,\qquad t\to+0,\label{heatex}
\end{equation}
Suppose that $\chi$ is a Laplace transform,
\begin{equation}
\chi (z)=\int_0^\infty dt\,e^{-tz} \tilde \chi (t)\,.\label{Lapl}
\end{equation}
Then, by substituting (\ref{Lapl}) in (\ref{Sb}) and using there the expansion (\ref{heatex}), one may show that
\begin{equation}
S_{\bf b}\left(\mathcal{D}\right)\sim\sum_{p=0}\Lambda^{4-2p}\chi_{2p}a_{2p}\left(D^{2}\right), \label{asymp}
\end{equation}
where
\begin{equation}
\chi_{2p}=\int_{0}^{\infty}dt\, t^{-2+p}\tilde \chi \left(t\right).
\end{equation}

To discuss the structure of expansion (\ref{asymp}) let us take the Dirac operator in the following simple but sufficiently general
form
\begin{equation}
D=i \gamma^\mu (\nabla^{\rm LC}_\mu +iA_\mu) + \gamma_5\phi \,,\label{exampD}
\end{equation}
where $\nabla^{\rm LC}$ is the Levi-Civita covariant derivative acting on spinors, $A_\mu$ is an abelian gauge field, and $\phi$ is a scalar field that
represents the Higgs field.  The chirality matrix $\gamma_5$ appears due to peculiarities of the Wick rotation, see \cite{doubling,doubling2} for the discussion. The presence of of $\gamma_5$ does not imply that $\phi$ is a pseudoscalar. 

To compute the heat kernel coefficients (see, e.g., \cite{Manual}), one first brings $D^2$ to the standard form
\begin{equation}
\mathcal{D}^2= -(\nabla^2+E)\,,\label{Lap}
\end{equation}
where \cite{KuLiVa}
\begin{equation}
E=-i\gamma^\mu \gamma_5 (\partial_{\mu}\phi) -\phi^2 -\tfrac 14 R +\tfrac i4 [\gamma^\mu,\gamma^\nu]F_{\mu\nu} \label{Ephi}
\end{equation}
with $F_{\mu\nu}\equiv \partial_\mu A_\nu -\partial_\nu A_\mu$. The connection $\nabla=\nabla^{\rm LC}+iA_\mu$ has the curvature
\begin{equation}
\Omega_{\mu\nu}\equiv [\nabla_\mu,\nabla_\nu] = iF_{\mu\nu} -\tfrac 14 \gamma^\sigma \gamma^\rho 
R_{\sigma\rho\mu\nu}\,.\label{Omega}
\end{equation}
For any operator of the form (\ref{Lap}), the heat kernel coefficients read
\begin{eqnarray}
&&a_0=\frac 1{16\pi^2} \int d^4x \sqrt {g} {\rm tr}\, (1)\nonumber\\
&&a_2=\frac 1{96\pi^2} \int d^4x \sqrt {g} {\rm tr}\, \big( 6E + R\big)\label{a024}\\
&&a_4=\frac 1{182\pi^2} \int d^4x \sqrt {g} {\rm tr}\, \big(\Omega_{\mu\nu}\Omega^{\mu\nu}+6E^2+2ER+R^2\mbox{--terms}\big)\nonumber
\end{eqnarray}
By using these formulas, one obtains the following expansion for the bosonic spectral action
\begin{eqnarray}
S_{\bf b}\left(\mathcal{D}\right)&\sim&\frac 1{48\pi^2} \int d^4x \sqrt {g} \big[ 12 \Lambda^4 \chi_0
+ \Lambda^2\chi_2(-2\phi^2-R) \label{Sbex}\\
&&\qquad + \chi_4 (2F_{\mu\nu}F^{\mu\nu} +6\phi^4+6(\partial_\mu\phi)^2+\mbox{higher curvature terms}) \big]
+\mathcal{O}(\Lambda^{-2})\nonumber
\end{eqnarray}
The action (\ref{Sbex}) contains the terms of all types that one expects from a bosonic action in four dimensions. By taking two constants,
$\chi_2$ and $\chi_4$ positive, we can ensure correct signs of the Einstein term, Maxwell term, the Higgs kinetic energy, mass and
selfcoupling, which is already a miracle. If one uses the standard model Dirac operator instead of
(\ref{exampD}) the similarities between the spectral action and the bosonic part of Standard Model become even more striking. Almost all terms come out correctly, except for a few problems:

\begin{enumerate}
\item One needs a not very natural renormalization procedure to get all $\Lambda^2$ terms right.
\item The Higgs mass appeared to be a bit too high, at about 170 GeV.
\item The Spectral Action predicts a unification point at some energy where all gauge coupling constants become equal -- the feature that contradicts the behavior of running couplings in the Standard Model.
\end{enumerate} 

I have nothing new to say about the first point above. The Higgs mass at 125 GeV may be very naturally achieved by including a new
scalar field, called $\sigma$, that has the meaning of fluctuating Majorana mass term. One may also consult the papers 
\cite{Stephan,DLM,CCvS} for various approaches to the Higgs mass and Standard Model within noncommutative geometry.

The expansion (\ref{asymp}) is an asymptotic $1/\Lambda$ expansion organized in accordance to the canonical mass dimension of the field polynomials. One assumes that the fields and their derivatives are small compared to the scale $\Lambda$.
There is no fundamental reason to terminate the expansion (\ref{asymp}) at the third term, as in (\ref{Sbex}). One can show \cite{vanSuijlekom:2011kc} that inclusion of higher order terms in the expansion of the Yang-Mills spectral action makes the theory superrenormalizable.
The $a_6$-term in the expansion of the Standard Model spectral action modifies the RG running and allows to achieve the unification of couplings \cite{DLFV}, thus solving one of the problems mentioned above.

\section{Very high energies}
The expansion (\ref{asymp}) for the spectral action is a low-energy asymptotic expansion. There are examples when it has a vanishing radius of
convergence. Therefore, it is desirable to construct another expansion for the spectral action that would be (i) convergent and (ii) valid for high energies. A suitable expansion of the spectral action was proposed in \cite{ILV2}, some mathematical aspects were developed further in
\cite{ILV3}, while physical consequences were discussed in \cite{KuLiVa}. Below we include a very brief overview of these works.

To simplify the discussion we take the cut-off function 
\begin{equation}
\chi (z)=e^{-z} \label{chiz}
\end{equation}
so that the spectral action becomes equal to the heat trace (\ref{heatex}) of $D^2$ with $t=1/\Lambda^2$. Let us assume, that 
\begin{equation}
\mathcal{D}^2=T+B\label{DTB}
\end{equation}
where $T$ is regarded as the ``main'' part, while $T$ is a perturbation. More precisely, we need $B$ to be relatively bounded with respect to
$T$. Consider the heat semigroup that is defined as $G(L,t)\equiv e^{-tL}$ for some operator $L$.
There is an expansion of the heat semigroup (called the Dyson-Phillips or Duhamel expansion):
\begin{equation}
G(T+B,t)=\sum_{n=0}^\infty G_n(t) \label{DPD}
\end{equation}
where
\begin{equation}
G_0(t)=G(T,t),\qquad G_{n+1}(t)=-\int_0^t G(T,t-s)BG_n(t) \label{iter}
\end{equation}
This expansion is trace-norm convergent and generates an expansion of the heat trace since $K(L,t)={\rm Tr}\, G(L,t)$. In applications,
it is essential to have a sufficiently simple operator $T$, so that the integrals in (\ref{iter}) may be computed. On an asymptotically flat
space with $\mathcal{D}$ given by (\ref{exampD}) it is natural to take $T=\mathcal{D}_0^2$, where $\mathcal{D}_0$ is the free Dirac operator on flat $\mathbb{R}^4$. Then one arrives at the Barvinsky-Vilkovisky expansion \cite{BV1}
\begin{eqnarray}
K(L,t)&\simeq& \frac 1{(4\pi t)^2} \int d^4x g^{\frac 12}\, {\rm tr}\, \left[ 1 + tP + t^2\bigl( 
R_{\mu\nu} f_1(-t\partial^2) R^{\mu\nu} + Rf_2(-t\partial^2)R \right.\nonumber\\
&& \left. + Pf_3(-t\partial^2)R +Pf_4(-t\partial^2) P + \Omega_{\mu\nu} f_5(-t\partial^2)\Omega^{\mu\nu}
\bigr) \right] + \dots \label{BVex}
\end{eqnarray}
where $P\equiv E+\tfrac 16 R$. The form-factors $f_{1,\dots,5}$ are expressed through the function 
\begin{equation}
h(z):=\int_0^1 d\alpha \,e^{-\alpha (1-\alpha)\,z}\,. \label{h}
\end{equation} 
For example,
\begin{equation*}
f_1(\xi)=\frac{h(\xi)-1+\tfrac 16 \xi}{\xi^2}
\end{equation*}
while other form-factors may be found in \cite{BV1}. 

In this expansion we assume that $E$, $R$ and $\Omega$ are small, but do not restrict their derivatives. In other words, (\ref{BVex}) is an expansion in fields, that is exact in the derivatives. This sounds better than the usual heat kernel expansion, but there is a price to pay:
the expansion in non-universal, depends in an essential way on the topology, and is much more complicated. In the low-energy limit 
$-t\partial^2\to 0$ on reproduces (\ref{a024}). Let us study the high energy limit, when the  $-t\partial^2$ is supposed to be large. Let us expand
the metric as $g_{\mu\nu}=\delta_{\mu\nu}+h_{\mu\nu}$, assuming that the fluctuations are gravitons, $\partial_\mu h_{\mu\nu}=0$ 
and $h_{\mu\mu}=0$. After some calculations, one obtains the quadratic in fluctuations part of (\ref{BVex}) keeping the leading terms in
the $-\partial^2/\Lambda^2\to \infty$ limit only:
\begin{equation}
K(\mathcal{D}^2,t=1/\Lambda^2)\simeq  
\frac {\Lambda^4}{(4\pi )^2} \int d^4x \left[ -\tfrac 32  h_{\mu\nu}h_{\mu\nu}
+8 \phi \frac 1{-\partial^2} \phi + 8 F_{\mu\nu} \frac 1{(-\partial^2)^2} F_{\mu\nu} \right] \label{expan}
\end{equation}

We see, that the expression (\ref{expan}) for the spectral action corresponding to the exponential cut-off (\ref{chiz}) depends
on zero or negative powers of the momenta. This means, that the propagators contain zero or positive powers of the momenta or that they are 
\emph{local}. In other words, high-momenta bosonic fields do not propagate!

It is interesting to compare the high energy expansion of the heat kernel for $\mathcal{D}^2$ to that of a generic Laplace type
operator. It appears that due to some miraculous cancellations the heat kernel for $\mathcal{D}^2$ falls off at large momenta faster than
that for a generic Laplacian.

If one takes a generic cut-off function instead of the exponent, the leading term in the high energy expansion of the bosonic spectral 
action differs by an overall coefficient from (\ref{expan}). All our conclusions remain unchanged.

\section{Conclusions and discussion}
Noncommutative geometry provides a unified framework for the description of both noncommutative and commutative manifolds. These scheme appears to be very restrictive, so that it excludes many field-theoretical models of particle physics. However, the standard model is
consistent with the axioms of noncommutative geometry and with the low energy expansion of the spectral action. We saw, that at high
energies the spectral action principle yields some universal predictions.  
The spectrum of bosonic particle should change dramatically at very high energies. In particular, this implies that there are no
high momentum gravitons, and the problem of perturbative nonrenormalizability of quantum gravity (as we know it) is no longer relevant.
Unfortunately, the calculation made in the previous section does not tell what the correct high energy gravity looks like. Therefore,
it is a bit too early to claim that the renormalization problem of quantum gravity is solved by the spectral action.

The characteristic scale $\Lambda$ that separates high energies is expected to be somewhere in the range of $10^{14}-10^{16}$GeV. 
There is a gap of a few orders of magnitude between $\Lambda$ and the Planck scale. It is quite possible therefore that the high energy
behavior of the spectral action will have some testable predictions, e.g. in Cosmology.

To conclude, I like to mention several relatively recent developments of the spectral action approach to particle physics. These are the
Pati-Salam unification \cite{PS}, the anomaly approach to the spectral action \cite{anomaly}, non-associative geometry approach to
the standard model \cite{na}, and a new zeta-function definition of the spectral action \cite{zeta}.

\ack
This contribution is based on joint papers with Agostino Devastato, Bruno Iochum, Max Kurkov, Cyril Levy, Fedele Lizzi and Carlos Valc\'arcel.
I am grateful to all of them for collaboration. This work was supported in part by the CNPq project 306208/2013-0 and by the FAPESP project 2012/00333-7.

\end{document}